\documentclass[12pt]{article}

\usepackage{scicite}
\usepackage{graphicx}

\usepackage{times}

\newenvironment{sciabstract}{%
\begin{quote} \bf}
{\end{quote}}

\title{The importance of shear on the collective charge transport in CDWs revealed by an XFEL source}

\author
{David Le Bolloc’h,$^{1\ast}$Ewen Bellec,$^{1,2}$ Darine Ghoneim,$^{1,3}$ Antoine Gallo-Frantz,$^{1}$\\ 
Pawel Wzietek,$^{1}$ Luc Ortega,$^{1}$ Anders Madsen,$^{3}$ Pierre Monceau,$^{4}$  Mathieu Chollet,$^{5}$\\
 Isabel Gonzales-Vallejo,$^{6}$ Vincent.L.R. Jacques,$^{1}$ and Aleksandr Sinchenko$^{1}$\\
 \\
\normalsize{$^{1}$Laboratoire de Physique des Solides, Université Paris-Saclay, CNRS, 91405 Orsay, France}\\
\normalsize{$^{2}$ESRF - The European Synchrotron, Grenoble, France}\\
\normalsize{$^{3}$European X-Ray Free-Electron Laser Facility, D-22869 Schenefeld, Germany}\\
\normalsize{$^{4}$Univ. Grenoble Alpes, CNRS, Grenoble INP, Institut Néel, 38000 Grenoble, France}\\
\normalsize{$^{5}$Linear Coherent Light Source, SLAC National Accelerator Laboratory, Menlo Park, CA 94025, USA}\\
\normalsize{$^{6}$Max-Born-Institute for Nonlinear Optics and Short Pulse Spectroscopy,}\\
\normalsize{ Max-Born-Straße 2A, 12489 Berlin, Germany}\\
\normalsize{$^\ast$Corresponding author. E-mail: david.le-bolloch@universite-paris-saclay.fr}
}

\date{}

\begin{document} 


\baselineskip24pt


\maketitle


\begin{sciabstract}
Charge transport in materials has an impact on a wide range of devices based on semiconductor, battery or superconductor technology. Charge transport in sliding Charge Density Waves (CDW) differs from all others in that the atomic lattice is directly involved in the transport process. To obtain an overall picture of the structural changes associated to the collective transport, the large coherent X-ray beam generated by an X-ray free-electron laser (XFEL) source was used. 
The CDW phase can be retrieved over the entire sample from diffracted intensities using a genetic algorithm. For currents below threshold, increasing shear deformation is observed in the central part of the sample while longitudinal deformation appears above threshold when shear relaxes. Shear thus precedes longitudinal deformation, with relaxation of one leading to the appearance of the other. Moreover, strain accumulates on surface steps in the sliding regime, demonstrating the strong pinning character of these surface discontinuities. The sliding process of nanometric CDW involves macroscopic sample dimensions.

\end{sciabstract}


\paragraph*{Introduction\\}
The Charge Density Wave (CDW) phase is dual in nature, displaying both an electronic and a structural phase transition. This metal to insulator transition occurs with the appearance of a gap at the Fermi level, as well as a Periodic Lattice Distortion (PLD) at the {\bf  2k$_F$} Fermi wave vector. The remarkable feature of the CDW phase is that it is highly sensitive to many external excitations, such as the temperature\cite{Moudden_prl90} or ultra-short laser pulses\cite{Singer2016,Jacques2016,Zong2019b,Yusupov2008,Gonzales2022,schm08}. The {\bf  2k$_F$} wave vector can even tilt for a few picoseconds after an ultra-short pulse in LaTe$_3$\cite{Kogar2020}. In the same family of compounds, the CDW is suppressed by small applied pressure\cite{Zocco_PRB2015} and is strongly modified by chemical pressures\cite{Ru2008}. Finally, a slight elongation of the atomic lattice along {\bf  2k$_F$} also  strongly increases the transition temperature, while a perpendicular elongation switches the CDW orientation\cite{straquadine_PRX2022,gallofrantz2023chargedensitywaves}.

However, the most interesting feature of an incommensurate CDW may well be its ability to {\it slide}. Under an external applied field exceeding a specific threshold value, a collective current appears. This charge transport  is unique because it is a pulsed and collective mode which occurs on macroscopic scale and leads to electronic noise oscillations, commonly referred to as Narrow-Band Noise\cite{PhysRevLett.45.43,Fleming_prl1979}. 
Furthermore, this collective current is directly linked to structural changes of the CDW. A correlation between the CDW structure and the collective current has been clearly observed in several CDW systems such as in TbTe$_3$\cite{Lebolloch_prb2016} and in blue bronze\cite{lebolloch_mdpi2023}. However,
the NbSe$_3$ system is the reference case. In this system, longitudinal deformations are observed, in particular a compression of the CDW near one electrode and an expansion near the other, but these only appear for applied currents above threshold\cite{PhysRevLett.70.845,PhysRevLett.80.5631}.
This phenomenon has been extensively studied and is interpreted to be a consequence of a phase slip phenomenon that allows the conversion of normal into condensed electrons in the CDW with wavefronts creation or suppression near the electrodes\cite{Ong:85om,PhysRevB.61.10640}.

A CDW submitted to a force exhibits longitudinal deformation, but also shear.  Shear deformation and its change were observed above threshold by X-ray topography\cite{Thorne_prl99} and below threshold by micro-diffraction \cite{bellec2019evidence}.
In the following, we show that the collective transport is not just related to longitudinal deformations localized near the electrodes, but involves shear that appears well before the threshold current, increases up to the threshold and then relaxes into the sliding state. The original approach proposed here, based on an XFEL study coupled with a genetic code, enables us to observe the balance between the two types of deformation, above and below threshold.

\begin{figure}
\centering
\includegraphics[scale=.5]{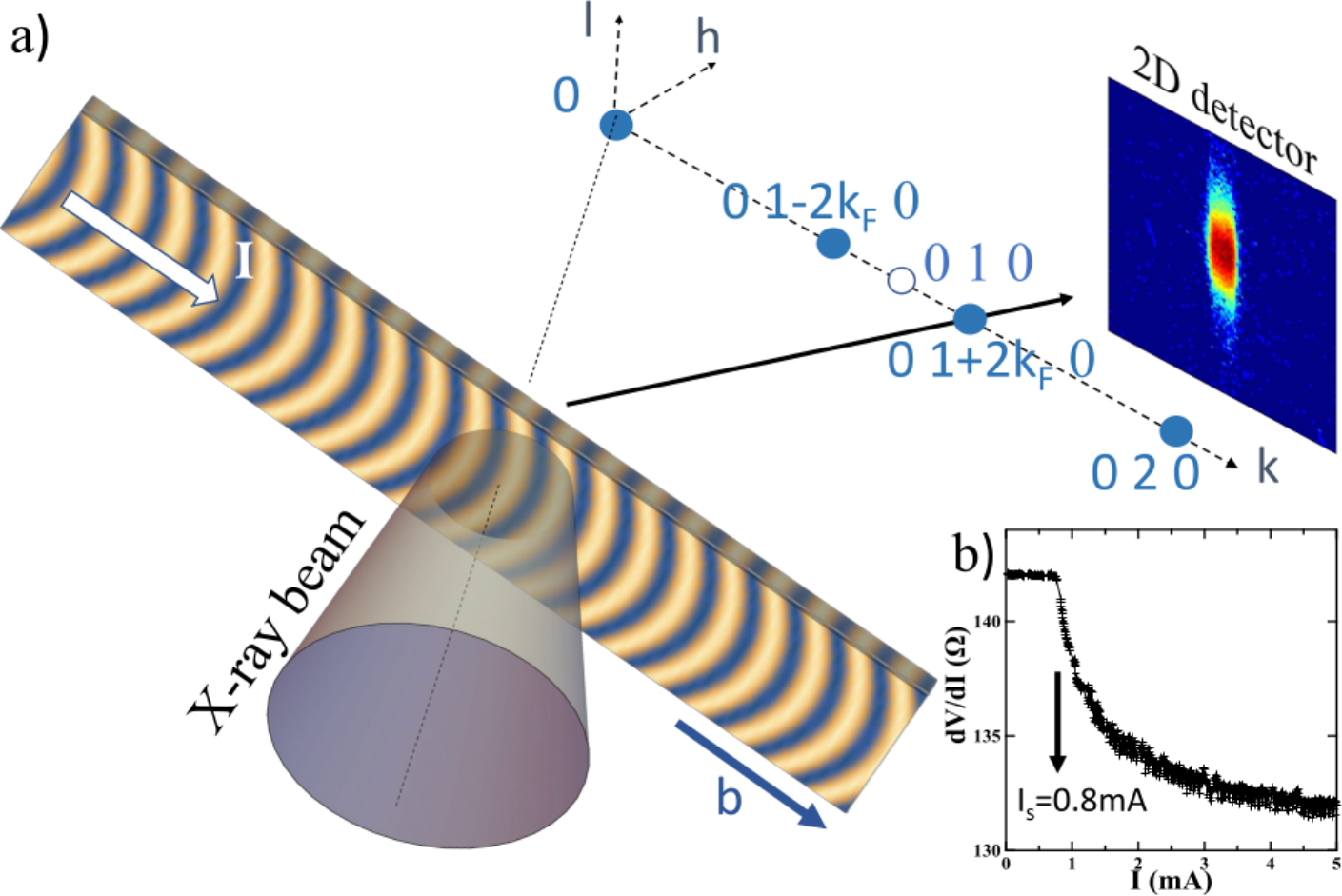}

{\bf Fig. 1.} {\bf Sketch of the experimental setup. (A)} The bended CDW in NbSe$_3$ probed by diffraction in wide angle configuration as well as the corresponding reciprocal lattice. The external current is applied along the {\bf  2k$_F$} wave vector, parallel to the b axis. {\bf (B)} NbSe$_3$ differential resistance measured during the experiment showing a threshold current equal to I$_S$=0.8mA.
\label{figure1}
\end{figure}

\paragraph*{Results\\}
This phenomenon has been observed by using the Linac Coherent Light Source at the XCS beamline\cite{xcs} that provides a large and fully coherent X-ray beam, which are the two necessary conditions for obtaining the CDW phase over a large area. The (0 1+q$_{s}$ 0) satellite reflection associated to the CDW in the NbSe$_3$ system has been probed by diffraction, with the incommensurate wave vector q$_{s}$= $0.243\pm$ 0.001 (see the sketch of the experimental setup in Fig. 1).  The important point for the following is that the measurement has been performed in the central part of the sample, far from the two electrodes, at more than 100$\mu m$ from them.  Integrated rocking curves  are shown in Fig.2 for several currents, below and above the threshold current I$_S$ (see Materials and Methods and \cite{Supplementary} for experimental details).  

Obviously, the satellite reflection is strongly disturbed with applied current. As the current increases until the threshold current I$_S$, the satellite peak spreads out along the vertical direction of the detector, corresponding to the {\bf  2k$_F$} transverse direction (q$_y$) (see transverse profile projection shown in Fig.2l). When the threshold current is exceeded, the diffraction profile contracts again, until it returns to a state closer in width to the initial one but with two maxima. Although the main deformation is transverse, a longitudinal deformation is also observed. 
Indeed, the longitudinal peak profile (Fig. 2m) shows a small shift near the threshold current I$_S$, and remains for greater currents. This longitudinal deformation corresponds to compression of the CDW period. As already mentioned in the introduction, the contraction/dilation that appears above threshold is known to be greater near the contacts, although it has also been observed far from the electrodes with a lower amplitude\cite{PhysRevLett.89.106404}, as is the case here. Finally, the longitudinal peak profile is slightly narrowing at the threshold  indicating an increase of the CDW longitudinal correlation length in the sliding regime (see Fig. S4c-d), in agreement with previous diffraction experiments\cite{PhysRevLett.89.106404}. At this point, it's important to note the obvious correlation between the two perpendicular deformations: below the threshold, only a transverse broadening is observed, without any change in the longitudinal direction. As  the threshold is approached and passed, the transverse width diminishes again while a longitudinal shift in peak position is observed.

\begin{figure*}
\centering
\includegraphics[scale= .5]{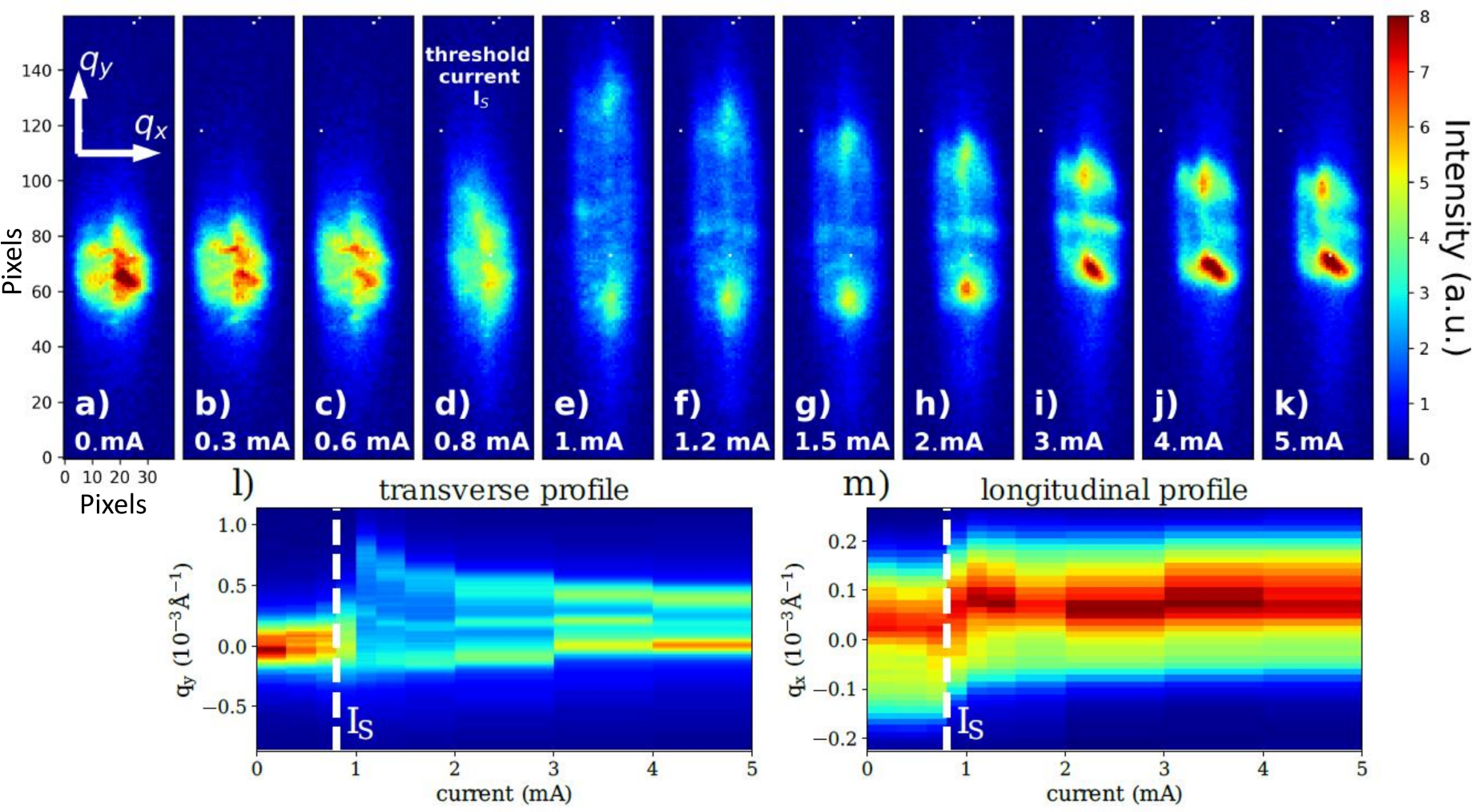}
{\bf Fig. 2.} {\bf Diffraction patterns versus current. (A)} to {\bf (K)} Integrated rocking curve of the (0 1+q$_{s}$ 0) satellite reflection associated to the CDW in NbSe$_3$ with increasing currents. The threshold current I$_S$=0.8mA is indicated in  {\bf (D)}. {(\bf L)} CDW peak projection along the transverse direction (q$_y$) for different currents.  {\bf (M)} CDW peak projection along the longitudinal direction (q$_x$).
\label{figure2}
\end{figure*}

The transverse lengthening of the satellite peak observed below threshold results mainly from the curvature of CDW wavefronts as already seen in a previous experiment based on scanning diffraction\cite{bellec2019evidence}. The bending of CDW wavefronts can be explained by considering the CDW as an elastic object strongly pinned by the electrodes and the lateral sample surfaces. The resulting phase is obtained using the image charge method, which gives, in 2D, and to first-order\cite{bellec2020}:
\begin{equation}
\phi(\vec{r})\propto-E \beta \cos(\pi \frac{x}{l_x})\cos(\pi \frac{y}{l_y})
\end{equation}
with the sample-size dependent parameter $\beta$:
$$\beta^{-1}=\frac{c_x^2}{l_x^2}+\frac{c_y^2}{l_y^2}$$ where $l_x$ is the distance between electrodes, $l_y$ the  sample width and $c_x,c_y$ the corresponding elastic coefficients. The sample thickness ($z$ direction) is very small compared to the other two and can be neglected. Details on the model are given in \cite{Supplementary} and the resulting topology displaying compression-dilatation close to electrodes and  wavefront curvature in the central part is shown in Fig. S6 and Fig. S7. Below threshold and away from the electrodes, the shear is dominant and the stress-strain tensor is reduced to the shear component. In that case, the stress $e$ and strain $\sigma$ can then be expressed as a function of the CDW phase, $e_{xy}=e_{yx}=-\frac 12\frac{1}{q_{s}}\frac{\partial\phi}{\partial y}$ and $\sigma_{xy}=\sigma_{yx}=q_{s}^2c_ye_{xy}$\cite{Feinberg}.
\begin{figure}
\centering
\includegraphics[scale= 0.7]{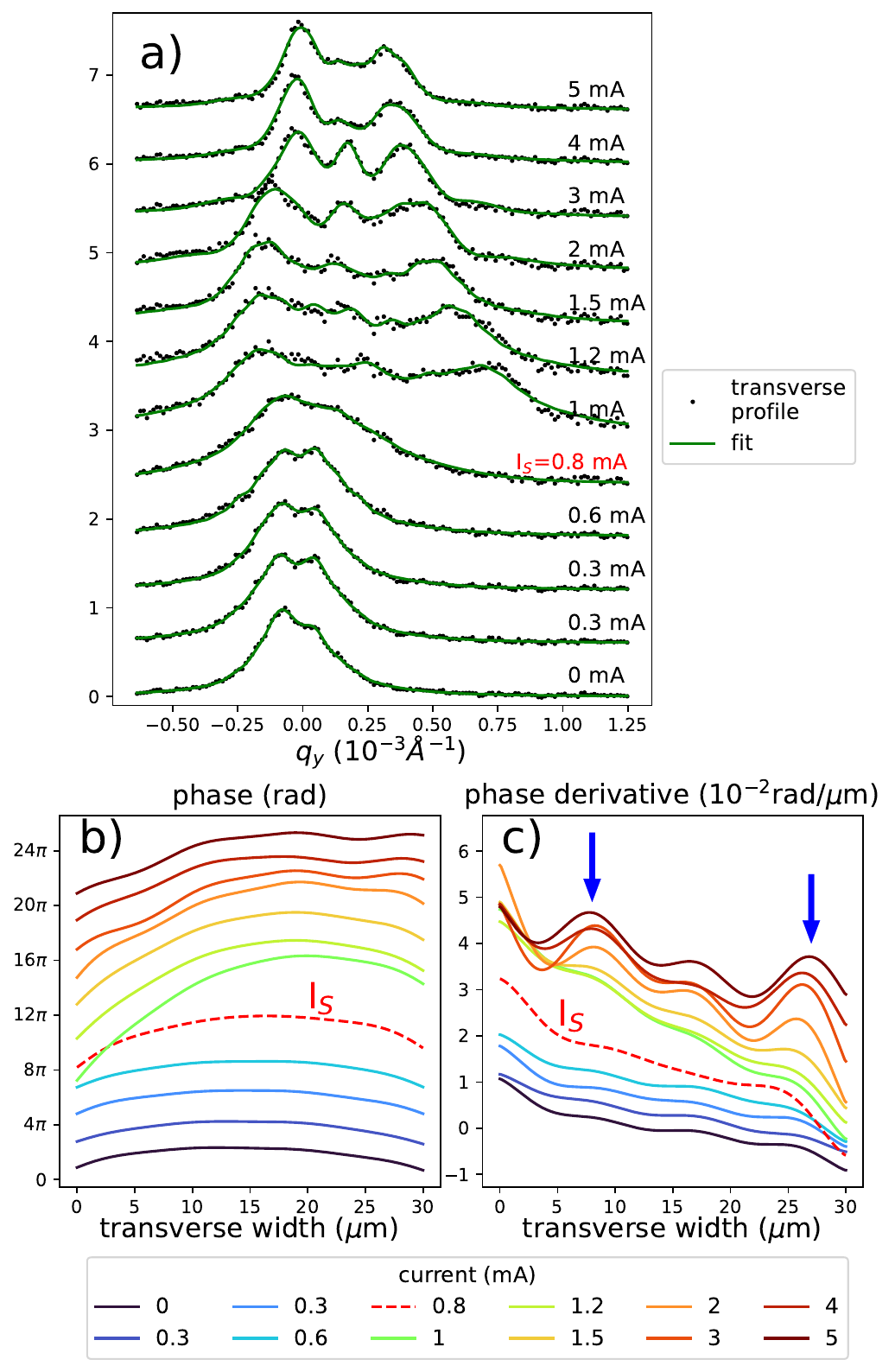}

{\bf Fig. 3.} {\bf CDW phase versus current.  (A)} Transverse profiles after vertical integration as well as the fit obtained by the genetic code. {\bf (B)} CDW phase versus current. Each curve is offset from the previous one by 2$\pi$. The phase at the threshold current I$_S$=0.8mA is shown as a red dashed line curve.  {\bf (C)} Phase derivative versus current (offset=0.3) with arrows indicating the two step positions.
\label{figure_fit}
\end{figure}

As a first step, the vertical diffraction profiles is interpreted by only considering shear, i.e. the $y$ dependence in Eq.1. The quadratic phase, $\phi(y) = \alpha y^2 + \beta y$, leading to curved CDW wavefronts, gives an expression for the diffracted intensity containing two Error functions (see \cite{Supplementary}). The overall behavior of the peak profiles versus current can be reasonably fitted by this simple model: the CDW curvature increases until threshold and decreases above (see Fig. S7 and Fig. S8b). However, a single CDW wavefront curvature from one edge of the sample to the other can not reproduce the observed asymmetric profiles. 

To go beyond this simple analytical model, a genetic code was used to obtain the CDW phase from diffracted intensities.
Phase retrieval from diffraction techniques still remains a challenge for conventional sources, and even more so for emerging XFEL sources, where the SASE process results in fluctuations of beam pointing. For example, ptychography methods\cite{Ptychography_prl2018} remain difficult to implement from those fluctuating sources. Here, an alternative retrieval method is employed that first consist in expanding the phase on basis of eigenfunctions and then appplying a genetic algorithm denoted Differential-Evolution (DE)\cite{storn_price} to fit all coefficients of the series. The DE method is a stochastic process which is not based on conventional gradient methods to solve the standard minimization problem, but probes large areas of configuration space and therefore, it can be used to solve even discontinuous or noisy cases. Iterative methods based on the DE algorithm are known to be very efficient for global optimization, despite the large number of adjustable parameters\cite{Royal-society-99,das2010de}. The algorithm is computationally intensive but one can take advantage of its inherent parallelism i.e., the $\chi^2$ calculation on a population of trial vectors can be done by independent threads.  The calculation of intensity profiles was coded in OpenCL and runs on a GPU, allowing to test several hundred thousand trial vectors per second. 

 The phase $\phi(y)$ has been expanded in terms a short Fourier series, whose first term corresponds to Eq.1.
 In addition,  a convolution with a resolution function containing a double Lorentzian has been used to take into account beam pointing errors and intensity fluctuations which enlarge the average beam width and extend the background. Finally, 14 adjustable parameters have to be considered.  
The best fit obtained after DE optimization is shown in Fig. 3A, together with the corresponding phase $\phi(y)$ in Fig. 3B, its derivative ${\partial\phi}/{\partial y}$ in Fig. 3C and the CDW state in real space in Fig. 4. The fit is excellent, taking into account the asymmetric profiles, with a steady evolution of coefficients (see Fig. S10).
Despite the large number of free parameters, and the few constraints imposed on each of them, the fit reveals relevant physical properties, in agreement with the previous analytical model (see Fig. S9). 

Before discussing these results, however, it is important to review the different ways of estimating the transverse deformation. Shear, which manifests itself as an elongation of the peak in the vertical direction, can be estimated using different approaches: either by the first coefficient of the Fourier series (the parameter C$_1$ in the Eq. 12 in the Suppl. Mat.), or by the averaged second derivative ($\langle\partial^2 \phi(y)/\partial y^2\rangle$), or by the quadratic term $\alpha$ in the phase expression used in the analytical model, or finally by the  transverse peak width characterized by its transverse standard deviation. All approaches yield  qualitative similar results and are presented in Fig. S11.
\begin{figure}
\centering
\includegraphics[scale= 0.5]{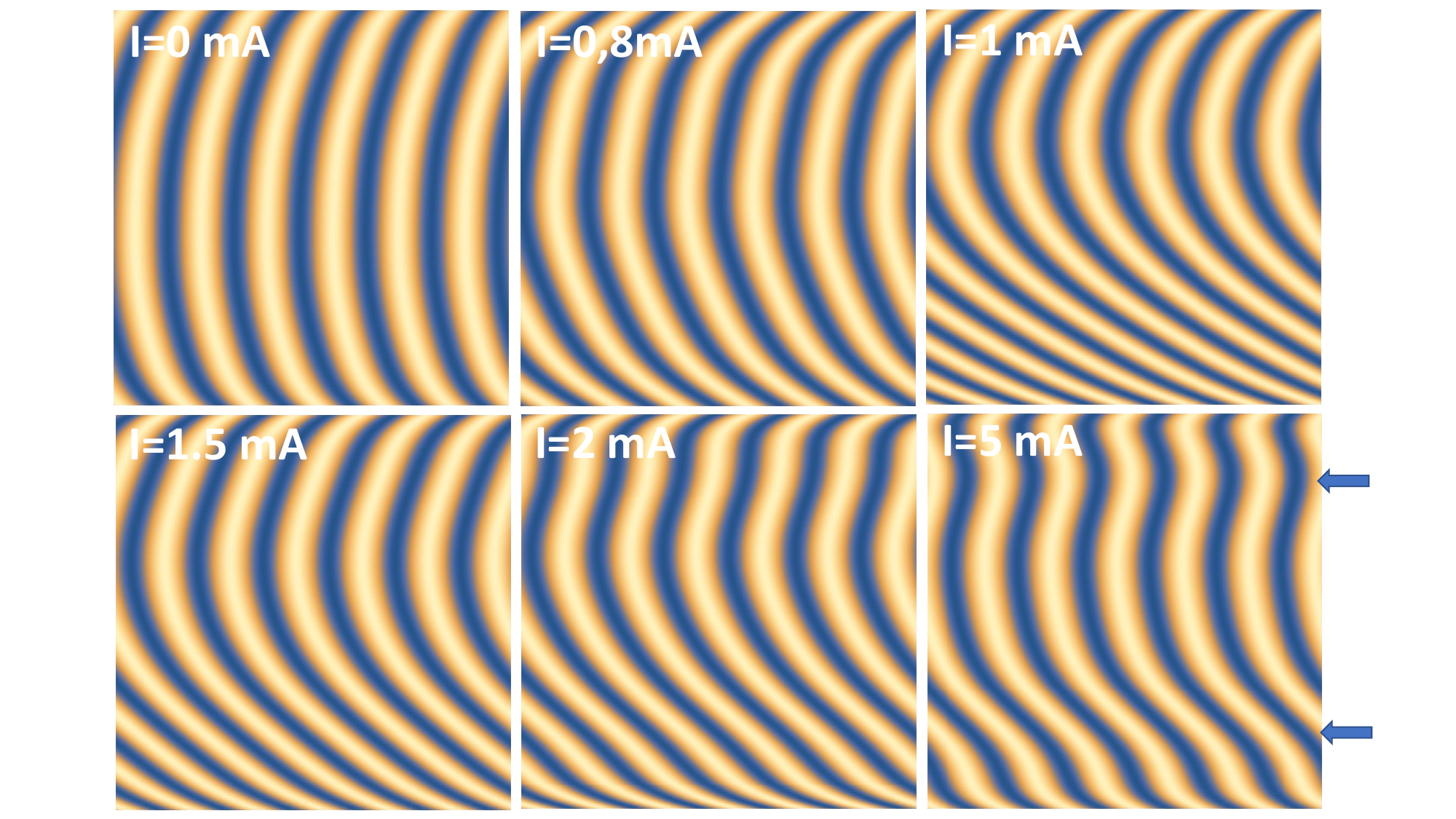}

{\bf Fig. 4. CDW state versus current.} 2D maps of the CDW state corresponding to 6 currents, below and above the threshold field (I$_S$=0.8mA) by using the CDW phase obtained from the fit. Wavefronts corresponding to a constant phase are represented in yellow. The two arrows show surface step position.
\label{phase_fit_2D}
\end{figure}

\paragraph*{Discussion\\}

First, let's look at the behavior of the phase below threshold. Note that the CDW is already curved at I=0mA (see Fig. 3B and 4), whereas we would expect flat wavefronts in the pristine state. This is due to a memory effect from earlier current injection performed to test the contact quality before measurement\cite{zettl1982onset}. 
This curvature then increases continuously with current, always exhibiting an almost quadratic phase, i.e. an almost linear derivative (see Fig. 3B and 3C). 
This effect is also observed from the standard deviation shown in Fig. 5  and from $\langle\partial^2 \phi(y)/\partial y^2\rangle$ or from the $C_1$ component. In the last case, the C$_1$ component is approximately proportional to the applied current, in a regime that can be considered as elastic (see Fig. S10A). This is in agreement with Eq.1 which considers an elastic CDW, strongly pinned by surfaces\cite{bellec2019evidence}.
The increasing curvature with current below threshold is in agreement with previous experiments based on scanning diffraction with a partially coherent microbeam\cite{bellec2020}. The similarity of the two results, obtained from two very different techniques, validates the DE code used here to retrieve the CDW phase.

\begin{figure}
\centering
\includegraphics[scale=.8]{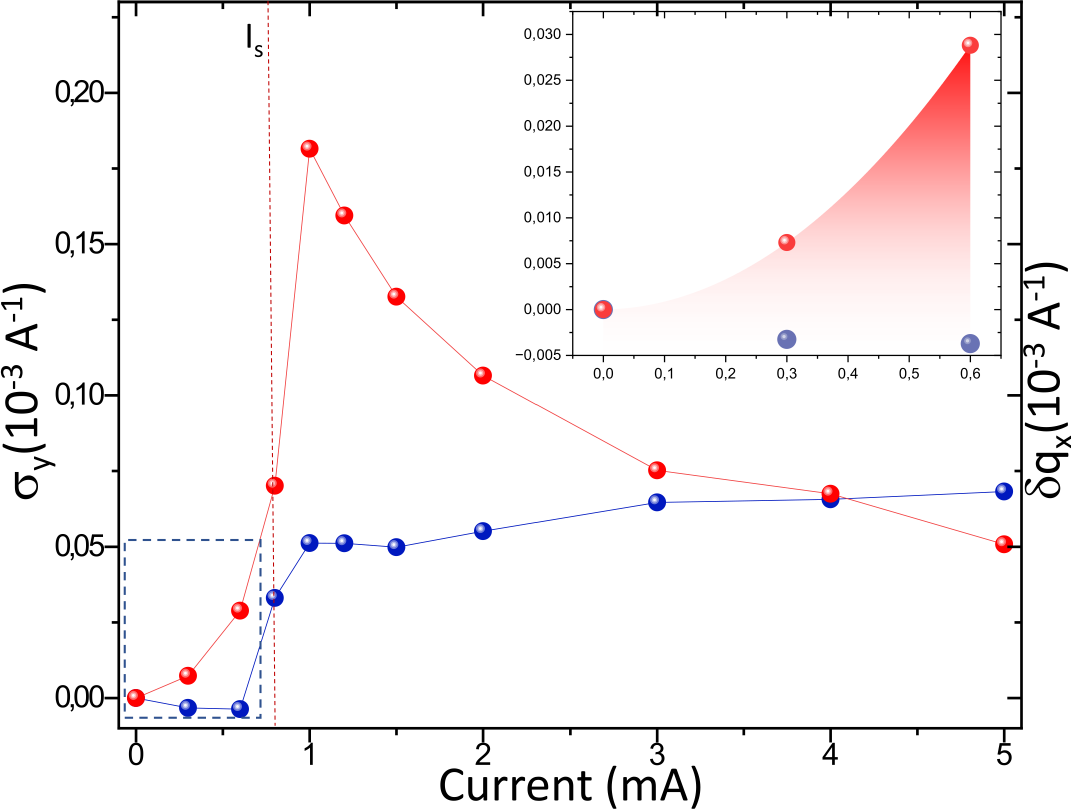}

{\bf Fig. 5. Longitudinal versus shear strain.} Comparison of the CDW peak transverse width $\sigma_y$ (shear) in red with the displacement of the longitudinal satellite position (compression) in blue for the different currents. The threshold current I$_S$ position is shown by the red dashed line. The inset is a zoom of the small currents below threshold, corresponding to the framed area (blue dashed region), highlighting the contrast between the exponential increase in shear and the longitudinal strain, which remains unchanged.
\label{shear_compression}
\end{figure}

The threshold current is associated with sudden CDW phase changes in both transverse and longitudinal directions, including a shear relaxation which had never been observed before. In the vicinity of the threshold field, the transverse phase tilts, becomes less and less sinusoidal and the curvature decreases (see Fig. 4). The phase tilt is responsible for the vertical shift of the satellite peak above the threshold as shown in Fig. 2A. The phase derivative, proportional to strain and stress, becomes less and less linear and its average slope decreases with current. A strong relaxation of the $C_1$ component is also observed above threshold (see Fig. S10A). This reduction in curvature above threshold  is also observed when using the quadratic model mentioned above (see Fig. S9).

Furthermore, the genetic code fit, while respecting the overall $\phi(y)$ curvature, reproduces the asymmetric profiles by adding local curvature variations. Indeed, the phase derivative displays two main maxima whose amplitude increases with current but whose position remains stable (see arrows in Fig. 2C). Obviously, these two maxima, appearing on top of the global curvature, emerge naturally from the fit without any constraint on its existence or position.
We speculate that the two pronounced maxima are the results of surface steps leading to local discontinuities in the wavefront. These can be seen in Fig. 4, where the phase extracted from the data is used to image the CDW in real space. The presence of surface steps is not surprising since NbSe$_3$ samples frequently exhibit surface steps as shown in the sample image (see Fig. S1).  

The observed shear relaxation (see Fig. 5) is the consequence of the dynamical character of the sliding regime. The strain seems to relax by a time-averaging effect. In the static and elastic regime, the static CDW bends until it reaches a maximum shear. The shear is then released due to  the periodic creation of CDW solitons that, once created and set in motion, are responsible for periodic oscillations of the electronic noise. A similar time-averaging effect has been observed by coherent x-ray diffraction, where the satellite reflection displays static speckles below threshold, which then disappear above once the CDW is set in motion\cite{Pinsolle_PRL2012}. 

The situation is reversed on steps, where no relaxation is observed above threshold, but instead an increase in the derivative with current is seen. The phase derivative on steps which appears increasingly strong with current, can be explained by a strong local pinning preventing sliding\cite{Feinberg,Thorne_prl99}. In the dynamical regime, the CDW slides everywhere except on steps, leading to flatter wavefronts between steps and locally enhanced shear strain on each step (see Fig. 2). The phenomenological elastic model  mentioned above can reproduce the behavior of the CDW between steps. The transition from the static state below threshold to the dynamic sliding regimere can be produced phenomenologically by decreasing the ratio between the two elastic constants $c_y/c_x$ (see Fig. S7).

The first conclusion of this paper is methodological. The phase retrieval method based on genetic code using the DE algorithm is an efficient approach to reconstruct phase objects from XFEL sources. It all depends on the choice of eigenfunction basis, which must be best suited to the physics under consideration. In our case, a Fourier basis has been chosen because the first order corresponds to the theoretical case, thus reducing the number of component.

Concerning the physics of incommensurate sliding CDW, we were able to disclose striking features. Firstly, shear precedes longitudinal strain below the threshold, and increases linearly until it displays a strong discontinuity at the threshold. Secondly, the dynamics of the sliding regime leads to time-averaged relaxation of the wavefront curvature, while on the contrary, strain accumulates on the surface steps due to the absence of sliding at these locations. The third point concerns the strong correlation existing between longitudinal and shear deformation (see Fig. 5). Longitudinal deformation appears when the transverse deformation drops and relaxes. The collective current in sliding CDW is linked to the two types of deformations, which are strongly correlated. The sliding process involves solitons nucleation in a macroscopic but finite  crystal, confined and driven by longitudinal {\it and} transverse sample boundaries.

\paragraph*{Materials and Methods}
\paragraph*{Diffraction experiment setup and details\\}
The incident 8 keV X-ray beam generated by the XFEL source is made of pulses of less than 50fs duration time at 120Hz repetition rate and was focused on a spot of about 30$\mu m$ in diameter at the sample position. The diffracted intensity was recorded with a 50$\mu$m$\times$50$\mu$m pixel size detector located 8m from the sample in the horizontal plane. 
The 39$\mu m$$\times$$3\mu m\times$$2.25mm$ NbSe$_3$ crystal was glued on a sapphire substrate and connected via two electrical contacts 500$\mu m$ apart and submitted to external $dc$ current (see Fig. S1 for the sample optical microscope image). The sample was cooled down to 100K on the XCS diffractometer, below the first CDW transition (Tc$_1$=145K). The (0 1+q$_{s}$ 0) with the incommensurate wave vector q$_{s}$= $0.243\pm 0.001 (\times \frac {2\pi} {b} $) was probed versus applied current. The threshold current, measured {\it in situ}, was equal to I$_S$ = 0.8 mA. The X-ray beam was localized in the central part of the sample, far from the two electrodes, at more than 100$\mu m$ from them (see the experimental setup in Fig. 1).
 Each image in Fig 2 corresponds to the sum over the entire rocking curve made of 40 angles, every 0.005° around the maximum peak intensity, with 480 images per angles, that is 19200 images in total. The rocking curve at I=0.3 mA was repeated twice to ensure measurement repeatability.
 \paragraph*{Genetic code\\}
  The genetic code used to retrieve the CDW phase fits the diffraction profiles for a population of trial vectors, at each iteration of the DE algorithm. This part was coded in OpenCL and run on the NVIDIA RTX 3060 GPU, enabling 700 thousand test configurations per second. 
The phase was expanded on a Fourier basis closely related to the elastic pinned model discussed previously,
 with the expected cosine function as the first term. A Lorentzian convolution has been used to take into account the direct beam profile. The fitting procedure was checked for optimum component number, stability and reproducibility (see Supplementary Materials Fig. S10). A tiny slope in the background is also used. 14 free parameters have been considered. \\

\noindent\textbf{Acknowledgments:}\\
\noindent\textbf{Funding:} Use of the Linac Coherent Light Source (LCLS), SLAC National Accelerator Laboratory, is supported by the
U.S. Department of Energy, Office of Science, Office of Basic Energy Sciences under Contract No. DEAC02-76SF00515.\\
\noindent\textbf{Author Contributions:}
      Conceptualization: DLB\\
      Sample preparation and transport measurement: AS\\
	Investigation: MC, EB, DG, AGF, PW, LO, AM, PM, IGV, VJ, AS\\
	Data treatement: EB, PW, DLB\\
	Supervision, writing—review and editing: DLB\\
	The authors would like to thank A. Rojo-Bravo for stimulating discussions.\\
\noindent\textbf{Competing interests}: All authors declare no competing interests.\\
\noindent\textbf{Data and Materials Availability:} All data needed to evaluate the conclusions in the paper are present in the paper and/or the Supplementary Materials.\\

\noindent\textbf{Supplementary materials} \\
Supplementary Text\\
Figs. S1 to S11\\
References (34-38)



\newpage

\clearpage

\end{document}